\documentclass[%
 aip,
 amsmath,
 amssymb,
 reprint,%
]{revtex4-1}

\usepackage{graphicx}
\usepackage{dcolumn}
\usepackage{bm}

\usepackage[utf8]{inputenc}
\usepackage[T1]{fontenc}
\usepackage{etoolbox}
\usepackage{ascmac,wrapfig,makeidx}
\usepackage{physics}
\usepackage[stable]{footmisc}
\usepackage{mathrsfs}
\usepackage{color}
\usepackage{comment}

\newcommand{\nn}{\nonumber}
\newcommand{\BE}{\begin{equation}}
\newcommand{\EE}{\end{equation}}
\newcommand{\be}{\begin{equation}}
\newcommand{\ee}{\end{equation}}
\newcommand{\BA}{\begin{eqnarray}}
\newcommand{\EA}{\end{eqnarray}}
\newcommand{\BAN}{\begin{eqnarray*}}
\newcommand{\EAN}{\end{eqnarray*}}
\newcommand{\BD}{\begin{description}}
\newcommand{\ED}{\end{description}}
\newcommand{\BEN}{\begin{enumerate}}
\newcommand{\EEN}{\end{enumerate}}

\newcommand{\BM}{\begin{minipage}}
\newcommand{\EM}{\end{minipage}}

\makeatletter
\def\@email#1#2{%
 \endgroup
 \patchcmd{\titleblock@produce}
  {\frontmatter@RRAPformat}
  {\frontmatter@RRAPformat{\produce@RRAP{*#1\href{mailto:#2}{#2}}}\frontmatter@RRAPformat}
  {}{}
}%
\makeatother

\begin{document}

\preprint{AIP/123-QED}

\title{Rigged Hilbert Space Approach for Non-Hermitian Systems with Positive Definite Metric}
\author{S.~Ohmori}
 \affiliation{Faculty of Science and Engineering, Waseda University, Tokyo 169-8555, Japan.}
\author{J.~Takahashi}%
\altaffiliation{These two authors contributed equally to this work.}
\affiliation{
Faculty of Science and Technology, Kochi University, Kochi 780-8072, Japan.
}%

\date{\today}

\begin{abstract}
We investigate Dirac's bra-ket formalism based on a rigged Hilbert space for a non-Hermitian quantum system with a positive-definite metric.
First, the rigged Hilbert space, characterized by positive-definite metric, is established.
With the aid of the nuclear spectral theorem for the obtained rigged Hilbert space, spectral expansions are shown for the bra-kets by the generalized eigenvectors of a quasi-Hermitian operator.
The spectral expansions are utilized to endow the complete bi-orthogonal system and the transformation theory between the Hermitian and non-Hermitian systems.
As an example of application, we show a specific description of our rigged Hilbert space treatment for some parity-time symmetrical quantum systems.
\end{abstract}

\maketitle


\section{Introduction}
\label{sec:1}

A rigged Hilbert space (RHS), also called a Gel'fand's triplet, was introduced mathematically to associate the distribution theory with the Hilbert space theory by I.~M.~Gel'fand and his collaborators \cite{Gelfand1964,Maurin1968}.
This space is characterized by the following triplet of topological vector spaces,
\begin{equation}
    \Phi \subset \mathcal{H} \subset \Phi^\prime,
    \label{eqn:O2-1}
\end{equation}
where $\mathcal{H}=(\mathcal{H}, \langle \cdot, \cdot \rangle_\mathcal{H})$ is a complex Hilbert space, $\Phi$ is a dense linear subspace of $\mathcal{H}$ that equips a topology $\tau_\Phi$ such that $(\Phi, \tau_\Phi)$ is a nuclear space, and $\Phi^\prime$ is a set of continuous linear functionals on $(\Phi,\tau_\Phi)$,
namely, a dual space of ($\Phi,\tau_{\Phi}$).
The inner product $\langle \cdot, \cdot \rangle_\Phi$ on $\Phi$ becomes  separately continuous on $(\Phi, \tau_\Phi)$, where $\langle \phi, \psi \rangle_\Phi \equiv \langle \phi, \psi \rangle_\mathcal{H}$ for $\phi, \psi \in \Phi$.
By the Rieszs' familiar theorem, $\mathcal{H}$ can be identified with its dual $\mathcal{H}'$, $\mathcal{H}=\mathcal{H}'$; then, the relation $\mathcal{H}'\subset \Phi^{\prime}$ holds.
Nowadays, the RHS is recognized as an underlying space for describing quantum mechanics 
\cite{Robert1966a,
Robert1966b,
Antoine1969a,
Antoine1969b,
Melsheimer1974a,
Melsheimer1974b,
Bohm1978,
Prigogine1996,
Bohm1998,
Antoniou1998,
Antoniou2003,
Gadella2003,
Madlid2004,
Madlid2005,
Antoine2009,
Liu2013,
Lars2019,
Antoine2021,
Fernandez2022}.
Specifically, accurate aesthetic formulations for quantum mechanics based on the RHS are shown for several systems, e.g., harmonic oscillator \cite{Bohm1978}, 1D rectangular barrier \cite{Madlid2004}, and non-Hermitian systems\cite{Fernandez2022}.
The RHS has also been focused on the fundamental study on how to characterize irreversibility found in non-equilibrium phenomena \cite{Prigogine1996,Bohm1998,Antoniou2003}.

For introducing the Dirac's bra-ket \cite{Dirac}, we consider the following triplet of topological vector spaces associating the RHS (\ref{eqn:O2-1}),
\begin{equation}
    \Phi \subset \mathcal{H} \subset \Phi^\times,
    \label{eqn:O2-1a}
\end{equation}
where $\Phi^\times$ is a set of continuous {\it anti}-linear functionals on $(\Phi,\tau_\Phi)$, namely, a function $f\in \Phi^{\times}$ that satisfies $f(a\varphi+b\phi)=\bar{a}f(\varphi)+\bar{b}f(\phi)$, where $a$ and $b$ are complex numbers with complex conjugates $\bar{a}$ and $\bar{b}$ and $\varphi, \phi \in \Phi$.
Given a quantum system, observables of the system that are identified with Hermitian operators in $\mathcal{H}$ define the space $\Phi$, and for the RHS, $\Phi^{\prime}$ and $\Phi^{\times}$  provide the bra and ket vectors as their elements, respectively.
The nuclear spectral theorem of the RHS shows the spectral expansions of the bra and ket vectors by the eigenvectors of the observables, and the Dirac's bra-ket formalism can be derived from the spectral expansions.

Recently, the RHS treatment is being used in studies on non-Hermitian quantum systems with a parity-time ($\mathcal{PT}$)-symmetry \cite{Lars2019,Fernandez2022}.
In a $\mathcal{PT}$-symmetrical system, a non-Hermitian Hamiltonian ${H}$ is assumed to satisfy the symmetric relation
\begin{equation}
{H} =\zeta {H} \zeta^{-1},
    \label{eqn:O1}
\end{equation}
where $\zeta$ is the operator that is given by the composition of the parity and time transformations \cite{Bender1998,Bender2007}.
%
%
L.~Laan applied the nuclear spectral theory of an RHS to quasi-Hermitian Hamiltonian system \cite{Lars2019}.
We believe that the RHS treatment is necessary to study the mathematical foundations for non-Hermitian quantum systems, such as a $\mathcal{PT}$-symmetrical non-Hermitian system.
However, studies of non-Hermitian systems from a mathematical viewpoint using RHS are insufficient now.

In this study, we discuss a non-Hermitian system with a symmetric structure based on RHS.
In particular, we focus on the bra-ket formalism of the system.
This paper is organized as follows.
An RHS associated with the positive-definite metric induced from a positive operator $\eta$, $\eta$-RHS,
is built in Section~\ref{sec:2}.
With the aid of the nuclear spectral theorem for the $\eta$-RHS, spectral expansions by the generalized eigenvectors of a $\eta$-quasi Hermitian operator for the bra and kets are obtained, and using the spectral expansions, the complete bi-orthogonal system that endows the transformation theory between the Hermitian and non-Hermitian systems is shown in Section~\ref{sec:3}.
The $\eta$-quasi Hermitian operator can be extended to the general operator on the bra-ket space while preserving its symmetry in order that
the observables in the non-Hermitian bra-ket space are considered.
In Secion~\ref{sec:4}, for the application of our $\eta$-RHS treatment to a physical model, the $\mathcal{PT}$-symmetrical system is focused on.
The conclusion is given in Section~\ref{sec:5}.

\section{$\eta$-Rigged Hilbert Space}
\label{sec:2}

We sketch out the flow of the definition of the bra-ket vectors
and their spectral expansions by generalized eigenvectors of an Hermitian operator based on the RHS (\ref{eqn:O2-1}) and (\ref{eqn:O2-1a}).
We consider an Hermitian operator $\eta$ in the Hilbert space $(\mathcal{H},\langle \cdot, \cdot \rangle_\mathcal{H})$ and
assume that $\eta$ is continuous on $(\Phi,\tau_\Phi)$ and $\eta \Phi \subset \Phi$.
The nuclear spectral theorem with respect to the RHS (\ref{eqn:O2-1}) endows the following properties \cite{Gelfand1964,Maurin1968}.
(i) $\eta$ has the set $\{ e_k(\xi)\}_{k=1}^{dim \hat{\mathcal{H}}(\xi)}$ of the generalized eigenvectors in $\Phi^{\prime}$ corresponding to generalized eigenvalues $\xi$, where each $\hat{\mathcal{H}}(\xi)$ is a Hilbert space, which composes of the direct integral that realizes $\mathcal{H}$, and $\xi$ goes through the spectra Sp$(\eta)\subset \mathbb{R}$ of $\eta$.
Here a linear functional $F$ in $\Phi^\prime$ is called a generalized eigenvector of $\eta$ corresponding to the eigenvalue $\xi$ when $F$ satisfies $F(\eta\phi)=\xi F(\phi)$ for every $\phi\in \Phi$\cite{Gelfand1964}.
(ii) For any $\varphi, \psi \in \Phi$, the relations
\begin{eqnarray}
    \langle \varphi, \psi \rangle_\mathcal{H} & = & \int_\mathbb{R}
     \sum_{k=1}^{dim \hat{\mathcal{H}}(\xi)} e_k(\xi)^*(\varphi)e_k(\xi)(\psi)
     d\nu (\xi),\nonumber \\
     \langle \varphi, \eta \psi \rangle_\mathcal{H} & = & \int_\mathbb{R}
     \sum_{k=1}^{dim \hat{\mathcal{H}}(\xi)} \xi~ e_k(\xi)^*(\varphi)e_k(\xi)(\psi) d\nu (\xi),
    \label{eqn:O2-2a}
\end{eqnarray}
are obtained, where $\nu(\xi)$ is a Borel measure on the spectrum of $\eta$.
(The symbol $*$ stands for the complex conjugate.)
Subsequently, the bra-ket notations can be constructed from (\ref{eqn:O2-2a}), as follows \cite{Madlid2005}.
Let $\varphi \in \Phi$.
We define a map
$\ket{\varphi}_\mathcal{H} : \Phi \rightarrow \mathbb{C}^1$, called a ket of $\varphi$ with respect to the $\mathcal{H}(=(\mathcal{H},{\langle \cdot, \cdot \rangle_\mathcal{H}}))$-system, by
$\ket{\varphi}_\mathcal{H}(\phi) \equiv  \langle \phi,\, \varphi\rangle_\mathcal{H}$ for $\phi \in \Phi$.
The bra of $\varphi$ with respect to the $\mathcal{H}$-system
is the map $\bra{\varphi}_\mathcal{H} : \Phi \rightarrow \mathbb{C}^1 $ of the complex conjugate of $\ket{\varphi}_\mathcal{H}$, namely, $\bra{\varphi}_{\mathcal{H}}(\phi)= \ket{\varphi}_{\mathcal{H}}^*(\phi)=(\ket{\varphi}_{\mathcal{H}}(\phi))^*=\langle \varphi,\, \phi\rangle_\mathcal{H}$.
$\bra{\varphi}_{\mathcal{H}}$ and $\ket{\varphi}_{\mathcal{H}}$ belong to $\Phi^{\prime}$ and $\Phi^{\times}$, respectively.
For simplicity, we assume $dim \, \hat{\mathcal{H}}(\xi)\equiv 1$ for any $\xi$ hereafter.
We denote the generalized eigenvector $e(\xi)$ by $\bra{\xi}_{\mathcal{H}}$ and $e(\xi)(\varphi)$ by $\braket{\xi}{\varphi}_{\mathcal{H}}$ ($\varphi \in \Phi$), respectively.
Additionally, we denote
$e(\xi)^*$ and $e(\xi)^*(\varphi)$ by $\ket {\xi}_{\mathcal{H}}$ and $\braket{\varphi}{\xi}_{\mathcal{H}}$, respectively. (See Appendix \ref{sec:6-1}  for information on the notations used in this paper.)
Using these notations, the relations in (\ref{eqn:O2-2a}) are represented as
\begin{eqnarray}
    \langle \phi, \varphi\rangle_\mathcal{H} &=& \int_\mathbb{R} \braket{\phi}{\xi}_{\mathcal{H}}\braket{\xi}{\varphi}_{\mathcal{H}} d\nu (\xi),
    \nonumber \\
    \langle \phi, \eta\varphi\rangle_\mathcal{H} &=& \int_\mathbb{R} \xi\braket{\phi}{\xi}_{\mathcal{H}}\braket{\xi}{\varphi}_{\mathcal{H}} d\nu (\xi)
    \label{eqn:O2-2b}
\end{eqnarray}
for any $\varphi,\phi\in\Phi$.
They endow the following spectral expansions for the bra $\bra{\varphi}_\mathcal{H}$ and ket $\ket{\varphi}_\mathcal{H}$ vectors in $\Phi^\prime$ and $\Phi^\times$ by the generalized eigenvectors $\{\bra{\xi}_{\mathcal{H}}\}$ and $\{\ket{\xi}_{\mathcal{H}}\}$ of $\eta$, respectively: for $\varphi\in\Phi$,
\begin{eqnarray}
    \bra{\varphi}_{\mathcal{H}} & = & \int_\mathbb{R} \braket{\varphi}{\xi}_{\mathcal{H}} \bra{\xi}_{\mathcal{H}} d\nu (\xi), \nn\\
    \bra{\eta\varphi}_{\mathcal{H}} & = & \int_\mathbb{R} \xi \braket{\varphi}{\xi}_{\mathcal{H}} \bra{\xi}_{\mathcal{H}} d\nu (\xi),
    \label{eqn:O2-3a}
\end{eqnarray}
and
\begin{eqnarray}
    \ket{\varphi}_{\mathcal{H}} & = &  \int_\mathbb{R} \braket{\xi}{\varphi}_\mathcal{H} \ket{\xi}_{\mathcal{H}} d\nu (\xi), \notag\\
    \ket{\eta\varphi}_{\mathcal{H}} & = & \int_\mathbb{R} \xi \braket{\xi}{\varphi}_{\mathcal{H}} \ket{\xi}_{\mathcal{H}} d\nu (\xi).
    \label{eqn:O2-3b}
\end{eqnarray}
In literature, the spectral expansions (\ref{eqn:O2-3a}) and (\ref{eqn:O2-3b}) are represented sometimes in the form
\begin{eqnarray}
    \ket{\varphi}_{\mathcal{H}} =
    \sum_{\xi_n \in Sp(\eta)} \braket{\xi _n}{\varphi}_\mathcal{H} \ket{\xi _n}_{\mathcal{H}}
    +
    \int_{\xi \in Sp(\eta)} \braket{\xi}{\varphi}_\mathcal{H} \ket{\xi}_{\mathcal{H}} d\nu (\xi)
\nn
\end{eqnarray}
where the sum is taken over the discrete spectrum and integral over the continuous spectrum.
Note that when we set a bra vector $\bra{\varphi}_\eta$ $(\varphi \in \Phi)$ by a map in $\Phi^{\prime}$ satisfying $\bra{\varphi}_\eta (\phi)
=\langle \eta\varphi, \, \phi\rangle_\mathcal{H}$ for any $\phi \in \Phi$
and a ket vector $\ket{\varphi}_\eta$ in $\Phi^{\times}$ as the complex conjugate of the bra vector $\bra{\varphi}_\eta$, $\ket{\varphi}_\eta(\phi)
= \langle \phi,\, \eta\varphi\rangle_\mathcal{H}$,
the relations
\BE
    \bra{\varphi}_\eta =\bra{\eta \varphi}_\mathcal{H}
    \textrm{ and }
    \ket{\varphi}_\eta =\ket{\eta \varphi}_\mathcal{H}
    \label{eqn:O2-4}
\EE
are satisfied.

Now, let $\eta$ be a positive operator on the Hilbert space $(\mathcal{H},\langle \cdot, \cdot \rangle_\mathcal{H})$
such that $\eta$ is continuous on $(\Phi,\tau_{\Phi})$ and satisfies $\eta \Phi \subset \Phi$.
Here, a linear bounded operator $\eta$ is said to be positive
if it satisfies $\langle \phi, \eta \phi\rangle_\mathcal{H} \geq 0$ for any $\phi \in H$.
Based on the positive operator $\eta$, we establish
another RHS as follows.
First, we introduce an inner product by
\begin{equation}
    \langle \phi, \psi \rangle_\eta
    = \langle \phi, \eta \psi \rangle_\mathcal{H} ~~~(\phi,\psi \in \mathcal{H}).
    \label{eqn:O2-6}
\end{equation}
The inner product (\ref{eqn:O2-6}) defines a new metric called the positive-definite metric, with respect to $\eta$, and considers the vector space $\mathcal{H}$ to be the  pre-Hilbert space $\mathcal{H}_\eta=(\mathcal{H},\langle \cdot, \cdot \rangle_\eta)$.
For $\mathcal{H}_\eta$, we denote its completion by $\widetilde{\mathcal{H}_\eta}=(\widetilde{\mathcal{H}_\eta},\widetilde{\langle \cdot, \cdot \rangle_\eta})$.
Then, it is shown that $\Phi$ is a dense linear subspace of $\widetilde{\mathcal{H}_\eta}$ and the inner product $\widetilde{\langle \cdot, \cdot \rangle_{\eta\Phi}}$, which is the restriction of $\widetilde{\langle \cdot, \cdot \rangle_\eta}$ to $\Phi$, is separately continuous on the nuclear space $(\Phi,\tau_\Phi)$.
Therefore, the triplet of the topological vector spaces
\begin{equation}
    \Phi \subset \widetilde{\mathcal{H}_\eta} \subset \Phi^\prime
    \label{eqn:O2-7}
\end{equation}
becomes an RHS,
where $\Phi^\prime$ is a set of continuous linear functionals on $(\Phi,\tau_\Phi)$.
Hereafter, we call (\ref{eqn:O2-7}) the $\eta$-RHS.
The bra and ket vectors with respect to the $\widetilde{\mathcal{H}_\eta}$-system can be constructed for $\eta$-RHS as well as for the original RHS (\ref{eqn:O2-1}), and they coincide with $\bra{\varphi}_\eta$ and $\ket{\varphi}_\eta$ defined in (\ref{eqn:O2-4}).

We consider an extension of $\eta$ to  $\Phi^{\prime}$ and $\Phi^{\times}$.
Because $\eta$ is continuous on $(\Phi, \tau_\Phi)$, we can define mappings
$\hat{\eta}^{\prime}:\Phi' \rightarrow \Phi'$ and $\hat{\eta}^{\times}:\Phi^{\times} \rightarrow \Phi^{\times}$
by $(\hat{\eta}^j (f))(\phi):=f(\eta (\phi))$, where $\phi\in\Phi,~f\in \Phi^j,~j=\prime,\times$.
These mappings can be combined into one operator $\hat{\eta} : \Phi^{\prime} \cup \Phi^{\times}\to\Phi^{\prime} \cup \Phi^{\times}$,
where
\BE
(\hat{\eta} (f))(\phi)=f(\eta (\phi))
    \label{eqn:O2-8}
\EE
for $f\in \Phi^{\prime} \cup \Phi^{\times}$ and $\phi\in\Phi$.
When $f\in \Phi^{j}$, we have $\hat{\eta} (f) \in \Phi^{j}$ ($j=\prime,\times$), and the relation
$\hat{\eta} (f^*)={\hat{\eta} ({f})}^*$ is satisfied.
Note that $\Phi^{\prime} \cap \Phi^{\times}=\{\hat{0}\}$ ($\hat{0}$ stands for the zero-valued functional on $\Phi$).
Moreover,
\BE
    \hat {\eta} \ket{\varphi}_\mathcal{H}
    =\ket{\varphi}_\eta, ~
    \bra{\varphi}_\mathcal{H} \hat {\eta}
    =\bra{\varphi}_\eta,
    \label{eqn:O2-9}
\EE
are obtained, where $ \bra{\varphi}_\mathcal{H} \hat {\eta}$ denotes $\hat {\eta} (\bra{\varphi}_\mathcal{H})$.
(For the derivation of the relation (\ref{eqn:O2-9}), see Appendix \ref{sec:A-2-2}.)
(\ref{eqn:O2-9}) shows that the operator $\hat{\eta}$ extended from $\eta$ transforms the bra and ket vectors of $\mathcal{H}$-system into those of the $\widetilde{\mathcal{H}_\eta}$-system.

When a positive operator $\eta$ on $\mathcal{H}$ is invertible,
$\widetilde{\mathcal{H}_\eta}=\mathcal{H}_\eta=(\mathcal{H}, \langle \cdot, \cdot \rangle_\eta)$.
$\langle \cdot, \cdot \rangle_\eta$ produces the equivalent norm of $\langle \cdot, \cdot \rangle_\mathcal{H}$.
The operator $\widehat{\eta^{-1}} : \Phi^{\prime} \cup \Phi^{\times}\to\Phi^{\prime} \cup \Phi^{\times}$ extended from the inverse $\eta^{-1}$ of $\eta$ can be defined by
\BE
(\widehat{\eta^{-1}} (f))(\phi)=f(\eta^{-1} (\phi))
    \label{eqn:O2-10}
\EE
for $f\in \Phi^{\prime} \cup \Phi^{\times}$ and $\phi\in\Phi$.
It follows from (\ref{eqn:O2-8}) and (\ref{eqn:O2-10}) that $\widehat{\eta^{-1}}$ is the inverse of $\hat{\eta}$, that is,
\BE
\hat{\eta}^{-1}=\widehat{\eta^{-1}} \nn,
\EE
and the following inverse relations to (\ref{eqn:O2-9}) are obtained:
\BE
    \hat {\eta}^{-1} \ket{\varphi}_\eta
    =\ket{\varphi}_\mathcal{H}, ~
    \bra{\varphi}_\eta \hat {\eta}^{-1}
    =\bra{\varphi}_\mathcal{H}.
    \label{eqn:O2-11}
\EE
%


\section{Bra-ket formalism for non-Hermitian
systems with positive-definite metric}
\label{sec:3}

%
%
%
%
%
%

\subsection{Spectral expansion}
\label{sec:3-1}

We consider a $\eta$-quasi Hermitian operator $A:\mathcal{D}(A)\to H$ from its domain $\mathcal{D}(A)$ in a Hilbert space $\mathcal{H}=(\mathcal{H}, \langle \cdot, \cdot \rangle_\mathcal{H})$ for a positive invertible operator $\eta$ \cite{Dieudonne1961,Mos2010,Antoine2013}.
Namely, $A$ satisfies
\BE
    A^{\dagger}= {\eta}
    A{\eta}^{-1},
    \label{eqn:O3-00}
\EE
where $A^{\dagger}$ is the adjoint of $A$
and $\eta$ becomes an intertwining operator for $A$ and $A^\dagger$.
Setting the  $\eta$-RHS (\ref{eqn:O2-7}) for $\mathcal{H}$ and $\eta$,
we suppose that $A\Phi \subset \Phi$, $\eta \Phi=\Phi$ and $A$ is continuous on $(\Phi,\tau_\Phi)$.
As $\eta$ is invertible, $\widetilde{\mathcal{H}_\eta}=\mathcal{H}_\eta=(\mathcal{H}, \langle \cdot, \cdot \rangle_\eta)$;
$\langle \cdot, \cdot \rangle_\eta$ produces the equivalent norm of $\langle \cdot, \cdot \rangle_\mathcal{H}$,
and hence the $\eta$-RHS becomes the following triplet based on ${\mathcal{H}_\eta}$:
\begin{equation}
    \Phi \subset {\mathcal{H}_\eta} \subset \Phi^\prime.
    \label{eqn:O3-0}
\end{equation}
The nuclear spectral theorem based on $\eta$-RHS (\ref{eqn:O3-0})
can be applied to $A$ because $A$ is an Hermitian operator in the Hilbert space $\mathcal{H}_\eta$.
Then, we have the following representations, as shown in Section~\ref{sec:2}, for $\eta$:
for any $\varphi,\phi\in\Phi$,
\begin{eqnarray}
    \langle \phi, \varphi\rangle_\eta &=& \int_\mathbb{R} \braket{\phi}{\lambda}_{\eta}\braket{\lambda}{\varphi}_{\eta} d\mu (\lambda),
    \label{eqn:O3-2a}
\end{eqnarray}
and
\begin{eqnarray}
    \langle \phi, A\varphi\rangle_\eta &=& \int_\mathbb{R} \lambda_A\braket{\phi}{\lambda}_{\eta}\braket{\lambda}{\varphi}_{\eta} d\mu (\lambda).
    \label{eqn:O3-2b}
\end{eqnarray}
Here, $\mu(\lambda)$ is a Borel measure on the spectrum of $A$,
$\bra{\lambda}_\eta$ and $\ket{\lambda}_\eta (\equiv \bra{\lambda}_\eta^*)$ are the generalized eigenvectors for $A$, which satisfy the following eigen equations:
\BA
    \bra{\lambda}_\eta(A\phi)=\lambda_A\bra{\lambda}_\eta(\phi)
    ,~
    \ket{\lambda}_\eta(A\phi)=\lambda_A\ket{\lambda}_\eta(\phi)
    \label{eqn:O3-2d}
\EA
for any $\phi \in \Phi$.
From (\ref{eqn:O3-2a})-(\ref{eqn:O3-2b}),  we obtain the following relations relative to $A$:
for $\varphi\in\Phi$,
\begin{eqnarray}
    \bra{\varphi}_{\eta} &=& \int_\mathbb{R} \braket{\varphi}{\lambda}_{\eta} \bra{\lambda}_{\eta} d\mu (\lambda),\nn\\
    \bra{A\varphi}_{\eta} &=& \int_\mathbb{R} \lambda_A \braket{\varphi}{\lambda}_{\eta} \bra{\lambda}_{\eta} d\mu (\lambda),
    \label{eqn:O3-3b}
\end{eqnarray}
and
\begin{eqnarray}
    \ket{\varphi}_{\eta} &=& \int_\mathbb{R} \braket{\lambda}{\varphi}_{\eta} \ket{\lambda}_{\eta} d\mu (\lambda), \nn\\
    \ket{A\varphi}_{\eta} &=& \int_\mathbb{R} \lambda_A \braket{\lambda}{\varphi}_{\eta} \ket{\lambda}_{\eta} d\mu (\lambda).
    \label{eqn:O3-3a}
\end{eqnarray}
(\ref{eqn:O3-3b}) and (\ref{eqn:O3-3a}) show the spectral expansions by the generalized eigenvectors of $A$ for the bra and ket in the $\mathcal{H}_\eta$-system, respectively.
In the $\mathcal{H}$-system the spectral expansions in terms of the bra $\bra{\varphi}_\mathcal{H}$ and ket $\ket{\varphi}_\mathcal{H}$ are obtained using the transformations (\ref{eqn:O2-4}), (\ref{eqn:O2-9}), and (\ref{eqn:O2-11}):
\BA
    \bra{\varphi}_\mathcal{H}
    &=& \int_\mathbb{R} \braket{\eta^{-1}\varphi}{\lambda}_\eta \bra{\lambda}_\eta d\mu(\lambda) \nn\\
    &=&\int_\mathbb{R}\bra{\varphi}_\mathcal{H}\hat{\eta}^{-1}\ket{\lambda}_\eta \bra{\lambda}_\eta d\mu(\lambda), \nn\\
    \ket{\varphi}_\mathcal{H}
    &=& \int_\mathbb{R} \braket{\lambda}{\eta^{-1}\varphi}_\eta \ket{\lambda}_\eta d\mu(\lambda) \nn\\
    &=& \int_\mathbb{R} \bra{\lambda}_\eta\hat{\eta}^{-1}\ket{\varphi}_\mathcal{H} \ket{\lambda}_\eta d\mu(\lambda).
    \label{eqn:O3-5a}
\EA
(The derivations of (\ref{eqn:O3-5a}); see Appendix \ref{sec:A-2-2}.)
Note that the bras $\bra{A\varphi}_\mathcal{H}$, $\bra{A^\dagger \varphi}_\mathcal{H}$ and kets $\ket{A\varphi}_\mathcal{H}$, $\ket{A^ \dagger \varphi}_\mathcal{H}$ are similarly formulated by (\ref{eqn:O3-2d})-(\ref{eqn:O3-5a}), as follows \cite{Lars2019}:
\BA
    \bra{A \varphi}_\mathcal{H}
     & = & \int_\mathbb{R} \lambda_A\bra{\varphi}_\mathcal{H}\ket{\lambda}_\eta \bra{\lambda}_\eta\hat{\eta}^{-1} d\mu(\lambda),\nn \\
    \ket{A\varphi}_\mathcal{H}
     & = & \int_\mathbb{R}\lambda_A \bra{\lambda}_\eta\ket{\varphi}_\mathcal{H} \hat{\eta}^{-1}\ket{\lambda}_\eta d\mu(\lambda),
   \label{eqn:O3-5c}
\EA
and
\BA
    \bra{A^\dagger \varphi}_\mathcal{H}
     & = & \int_\mathbb{R}\lambda_A\bra{\varphi}_\mathcal{H}\hat{\eta}^{-1}\ket{\lambda}_\eta \bra{\lambda}_\eta d\mu(\lambda),\nn \\
    \ket{A^\dagger \varphi}_\mathcal{H}
     & = &  \int_\mathbb{R}\lambda_A \bra{\lambda}_\eta\hat{\eta}^{-1}\ket{\varphi}_\mathcal{H} \ket{\lambda}_\eta d\mu(\lambda).
   \label{eqn:O3-5d}
\EA
Note that the spectral expansions for the bra-ket vectors obtained here reproduce those found in Ref. \cite{Lars2019}.

\subsection{Complete bi-orthogonal system}
\label{sec:3-2}

From the relations in (\ref{eqn:O3-5a}), we find the completion form
\BA
    I =  \int_\mathbb{R} \ket{\lambda}_\eta \bra{\lambda}_\eta\hat{\eta}^{-1}  d\mu(\lambda)
    =\int_\mathbb{R} \hat{\eta}^{-1} \ket{\lambda}_\eta \bra{\lambda}_\eta  d\mu(\lambda),
    \label{eqn:O3-7}
\EA
where $I$ is the identity for the bra $\bra{\cdot}_\mathcal{H}$ and ket $\ket{\cdot}_\mathcal{H}$.
Here, we introduce two operators $\bra{\lambda}_\mathcal{H}$ and $\ket{\lambda}_\mathcal{H}$ in $\Phi^{\prime}$ and $\Phi^{\times}$ as well as (\ref{eqn:O2-11}) by
\BE
    \bra{\lambda}_\mathcal{H}=\bra{\lambda}_\eta\hat{\eta}^{-1}
    \text{~and~}
    \ket{\lambda}_\mathcal{H}=\hat{\eta}^{-1}\ket{\lambda}_\eta,
    \label{eqn:O3-110}
\EE
and set $\varphi^{\eta}(\lambda)=\bra{\lambda}_\eta (\varphi)$
for $\bra{\lambda}_\eta$ and $\varphi \in \Phi$.
Using (\ref{eqn:O3-7}) and the notation (iv) in Appendix \ref{sec:6-1}, we have
\BA
    &&\int_\mathbb{R} \braket{\lambda}{\lambda^{\prime}}_\eta \varphi^{\eta}(\lambda ^{\prime}) d\mu(\lambda^{\prime}) \nn\\
    & = & \int_\mathbb{R} \bra{\lambda}_\eta \cdot \ket{\lambda^\prime}_\mathcal{H} \bra{\lambda^\prime}_\eta (\varphi)  d\mu(\lambda^{\prime})
    \nn \\
    & = & \int_\mathbb{R} \bra{\lambda}_\eta \cdot \hat{\eta}^{-1}\ket{\lambda^\prime}_\eta \bra{\lambda^\prime}_\eta \cdot \ket{\varphi}_\mathcal{H}  d\mu(\lambda^{\prime})
    \nn \\
     & = & \bra{\lambda}_\eta\cdot \ket{\varphi}_\mathcal{H}
     \nn \\
     & = & \varphi^{\eta}(\lambda).
    \label{eqn:O3-8}
\EA
Thus, the orthonormal relation is obtained as
\BE
    \braket{\lambda}{\lambda^{\prime}}_\eta
    =\bra{\lambda}_\eta \hat{\eta}^{-1}\ket{\lambda^\prime}_\eta
    =\delta(\lambda - \lambda^\prime),
    \label{eqn:O3-10}
\EE
where $\delta$ is Diracs' $\delta$-function
as the normalization factor of the eigenvectors of $A$.
Therefore, from (\ref{eqn:O3-7}), (\ref{eqn:O3-110}), and (\ref{eqn:O3-10}), it is confirmed that the pair $\{ \ket{\lambda}_\eta, \ket{\lambda}_\mathcal{H} \}$
composes of the complete bi-orthogonal system:
\BA
    I =  \int_\mathbb{R} \ket{\lambda}_\eta \bra{\lambda}_\mathcal{H}  d\mu(\lambda)
    =\int_\mathbb{R} \ket{\lambda}_\mathcal{H} \bra{\lambda}_\eta  d\mu(\lambda),
    \label{eqn:O3-11a}
\EA
and
\BE
    \bra{\lambda}_\mathcal{H}\cdot \ket{\lambda^{\prime}}_\eta
    =\bra{\lambda}_{\eta} \cdot\ket{\lambda^{\prime}}_\mathcal{H}
    =\delta(\lambda - \lambda^\prime).
    \label{eqn:O3-11b}
\EE
%
%

\subsection{Transformation theory}
\label{sec:3-4}

Set an Hermitian operator $B:\mathcal{D}(B)\rightarrow \mathcal{R}(B)$ in RHS (\ref{eqn:O2-1}), namely, $B$ is Hermitian in $(\mathcal{H},\langle \cdot, \,  \cdot\rangle_\mathcal{H})$
satisfying continuity on $(\Phi,\tau_\Phi)$ and  $B\Phi\subset\Phi$.
($B=\eta$ is possible.)
The nuclear spectral theorem based on RHS shows that each of the bra $\bra{\varphi}_\mathcal{H}$ and ket $\ket{\varphi}_\mathcal{H}$ is expanded in the $\mathcal{H}$-system by the generalized eigenvectors $\{\bra{\omega}_\mathcal{H}\}$ of $B$,
\begin{eqnarray}
    \bra{\varphi}_{\mathcal{H}} &=& \int_\mathbb{R} \braket{\varphi}{\omega}_{\mathcal{H}} \bra{\omega}_{\mathcal{H}} d\mu (\omega)
    , \nn\\
    \ket{\varphi}_{\mathcal{H}} &=& \int_\mathbb{R}  \braket{\omega}{\varphi}_{\mathcal{H}} \ket{\omega}_{\mathcal{H}} d\mu (\omega),
    \label{eqn:O3-4-1}
\end{eqnarray}
where $\mu(\omega)$ is a Borel measure on the spectrum of $B$ and $\braket{\omega}{\varphi}_{\mathcal{H}}=\bra{\omega}_\mathcal{H}(\varphi)$ and $\braket{\varphi}{\omega}_{\mathcal{H}}=\ket{\omega}_\mathcal{H}(\varphi)$.
As known from the RHS formulation \cite{Madlid2005}, the generalized eigenvectors $\{\bra{\omega}_\mathcal{H}\}$ composes of the complete orthonormal system for the $\mathcal{H}$-system:
\BA
\braket{\omega}{\omega^{\prime}}_\mathcal{H}=\delta (\omega-\omega^{\prime})
    \label{eqn:O3-4-2a}
\EA
and
\BA
\int_\mathbb{R}  \ket{\omega}_{\mathcal{H}}\bra{\omega}_\mathcal{H} d\mu (\omega)=I.
    \label{eqn:O3-4-2b}
\EA
The complete relations ({\ref{eqn:O3-4-2b}}) for $\{\ket{\omega}_\mathcal{H}\}$ and (\ref{eqn:O3-11a}) for $\{\ket{\lambda}_\mathcal{H},\ket{\lambda}_\eta\}$ tell us the procedure of the transformations of the bra-ket vectors of their representations.
For any $\varphi\in \Phi$, we obtain
\BA
    \braket{\omega}{\varphi}_\mathcal{H}
    & = & \bra{\omega}_\mathcal{H}\cdot \ket{\varphi}_\mathcal{H}\nn \\
    & = & \int_\mathbb{R} \bra{\omega}_\mathcal{H}\cdot \ket{\lambda}_\eta\bra{\lambda}_\mathcal{H}\cdot\ket{\varphi}_\mathcal{H}  d\mu (\lambda)\nn \\
    & = & \int_\mathbb{R} \braket{\omega}{\lambda}_\eta\braket{\lambda}{\varphi}_\mathcal{H}  d\mu (\lambda).
    \label{eqn:O3-4-3a}
\EA
The relation (\ref{eqn:O3-4-3a}) shows the transformation of the representations from $\bra{\lambda}_\mathcal{H}$ to $\bra{\omega}_\mathcal{H}$ by means of the transformation factor $\bra{\omega}_\mathcal{H}\cdot \ket{\lambda}_\eta=\braket{\omega}{\lambda}_\eta$.
Additionally, we have
\BA
    \braket{\omega}{\varphi}_\mathcal{H}
     &=&  \int_\mathbb{R} \bra{\omega}_\mathcal{H}\cdot \ket{\lambda}_\mathcal{H}\bra{\lambda}_\eta\cdot\ket{\varphi}_\mathcal{H}  d\mu (\lambda) \nn\\
     &=&  \int_\mathbb{R} \braket{\omega}{\lambda}_\mathcal{H}\braket{\lambda}{\varphi}_\eta  d\mu (\lambda),
    \label{eqn:O3-4-3b}
\EA
which endows the transformation from $\bra{\lambda}_\eta$ to $\bra{\omega}_\mathcal{H}$ by means of the transformation factor $\bra{\omega}_\mathcal{H}\cdot \ket{\lambda}_\mathcal{H}=\braket{\omega}{\lambda}_\mathcal{H}$.
Similarly, we obtain the other factors
$\braket{\lambda}{\omega}_\mathcal{H}$
and $\braket{\lambda}{\omega}_\eta$ that characterize the following transformation equations from $\bra{\omega}_\mathcal{H}$
to $\bra{\lambda}_\mathcal{H}$ and $\bra{\lambda}_\eta$, respectively:
\BA
    \braket{\lambda}{\varphi}_\mathcal{H}
     &=&  \int_\mathbb{R} \braket{\lambda}{\omega}_\mathcal{H}\braket{\omega}{\varphi}_\mathcal{H}  d\mu (\omega), \nn\\
    \braket{\lambda}{\varphi}_\eta
     &=&  \int_\mathbb{R} \braket{\lambda}{\omega}_\eta\braket{\omega}{\varphi}_\mathcal{H}  d\mu (\omega).
    \label{eqn:O3-4-3c}
\EA
Note that the following relations of the complex conjugates
are satisfied:
\BE
    \braket{\lambda}{\omega}_\eta ^*=\braket{\omega}{\lambda}_\eta
    , \quad
    \braket{\lambda}{\omega}_\mathcal{H} ^*=\braket{\omega}{\lambda}_\mathcal{H}.
    \label{eqn:O3-4-5}
\EE
(The derivation of (\ref{eqn:O3-4-5}), see Appendix \ref{sec:A-2-5}.)

We suppose that the transformation factors supply the Fourier transformation;
set $\braket{\omega}{\lambda}_\eta=\frac{1}{\sqrt{2\pi\hbar}}e^{-i\omega\lambda/\hbar}$ in (\ref{eqn:O3-4-3a}) such that $\lambda$ and $\omega$ are like the position and momentum operators in Hermitian quantum mechanics.
Then, $\braket{\lambda}{\omega}_\eta=\braket{\omega}{\lambda}_\eta ^*=\frac{1}{\sqrt{2\pi\hbar}}e^{i\omega\lambda/\hbar}$ from (\ref{eqn:O3-4-5}).
By the transformations in (\ref{eqn:O3-4-3c}),
\BA
\braket{\lambda}{\varphi}_\eta
     =  \frac{1}{\sqrt{2\pi \hbar}}\int_\mathbb{R} e^{i\omega\lambda/\hbar}\braket{\omega}{\varphi}_\mathcal{H}  d\mu (\omega)
     =\braket{\lambda}{\varphi}_\mathcal{H}
     \nn
\EA
is obtained for any $\varphi\in \Phi$.
Thus, $\bra{\lambda}_\eta=\bra{\lambda}_\mathcal{H}$, which shows that $\hat{\eta}$ is the identity.
Therefore, the transformation of the representations between the $\mathcal{H}$-system and $\mathcal{H}_\eta$-system is different from the Fourier transformation that is the typical transformation for Hermitian systems such as between the real space and the momentum space.

We note that when the transformation relations (\ref{eqn:O3-4-3a})-(\ref{eqn:O3-4-3c}) are given first,
the bi-orthogonal relations for $\{\ket{\lambda}_\mathcal{H},\ket{\lambda}_\eta\}$ can be derived;
for instance, by (\ref{eqn:O3-4-3b}),
\BA
    &&\braket{\lambda}{\varphi}_\mathcal{H} \nn\\
     & = & \int_\mathbb{R} \braket{\lambda}{\omega}_\mathcal{H}\braket{\omega}{\varphi}_\mathcal{H}  d\mu (\omega)\nn \\
      & = & \int_\mathbb{R} \braket{\lambda}{\omega}_\mathcal{H}
      \Big{(} \int_\mathbb{R} \braket{\omega}{\lambda^{\prime}}_\eta\braket{\lambda^{\prime}}{\varphi}_\mathcal{H}  d\mu (\lambda^{\prime})\Big{)}d\mu (\omega)\nn \\
      & = & \int_\mathbb{R}\int_\mathbb{R} \braket{\lambda}{\omega}_\mathcal{H}\braket{\omega}{\lambda^{\prime}}_\eta\braket{\lambda^{\prime}}{\varphi}_\mathcal{H}  d\mu (\omega)d\mu (\lambda^{\prime})\nn \\
      & = & \int_\mathbb{R}\bra{\lambda}_\mathcal{H}\cdot
      \int_\mathbb{R} \ket{\omega}_\mathcal{H}\bra{\omega}_\mathcal{H} d\mu (\omega)
      \cdot \ket{\lambda^{\prime}}_\eta\braket{\lambda^{\prime}}{\varphi}_\mathcal{H}  d\mu (\lambda^{\prime})\nn \\
      & = & \int_\mathbb{R} \braket{\lambda}{\lambda^{\prime}}_\eta  \braket{\lambda^{\prime}}{\varphi}_\mathcal{H} d\mu (\lambda^{\prime}),
    \label{eqn:O3-4-4}
\EA
which shows the relation (\ref{eqn:O3-11b}).

\subsection{Extensions}
\label{sec:3-3}

An $\eta$-quasi Hermitian operator $A$ can be extended to the set of bra-kets while retaining its symmetry.
The fact that $A$ is continuous on $(\Phi,\tau_\Phi)$ with $A\Phi \subset \Phi$ provides an operator
$\hat{A}$ on $\Phi^{\prime}\cup\Phi^{\times}$, given by
\BE
   (\hat {A} (f))(\phi):=f(A(\phi)),
    \label{eqn:O3-2-1}
\EE
for any $\phi\in\Phi$ and $f\in \Phi^\prime\cup \Phi^\times$.
From (\ref{eqn:O3-2d}), $\hat {A}$ satisfies the following eigenequations for the generalized eigenvectors $\{\bra{\lambda}_\eta\}$ and $\{\ket{\lambda}_\eta\}$ of $A$:
\BE
    \bra{\lambda}_\eta\hat{A} =\lambda_A\bra{\lambda}_\eta,~
    \hat{A}\ket{\lambda}_\eta=\lambda_A\ket{\lambda}_\eta,
    \label{eqn:O3-2-2}
\EE
where we denote $\hat{A}(\bra{\lambda}_\eta)$ by $\bra{\lambda}_\eta\hat{A}$.
The disjoint operator $A^\dagger$ can be also generalized to the operator $\hat{A}^{\dagger}$ on $\Phi^\prime\cup \Phi^\times$, where
\BE
    (\hat {A}^\dagger (f))(\phi):=f(A^\dagger(\phi))
    \label{eqn:O5-4}
\EE
for $\phi\in\Phi,~f\in \Phi^\prime\cup \Phi^\times$.
It is easy to check that the following eigenequations are obtained:
\BE
    \hat{A}^{\dagger}\bra{\lambda}_\mathcal{H} =\lambda_A \bra{\lambda}_\mathcal{H},~
    \hat{A}^{\dagger}\ket{\lambda}_\mathcal{H} =\lambda_A \ket{\lambda}_\mathcal{H}.
    \label{eqn:O5-7}
\EE
Regarding the symmetry between $\hat{A}$ and $\hat{A}^\dagger$ with respect to $\hat{\eta}$,
we can prove the relation
\BE
    \bra{\phi}{\hat A^{\dagger}}\ket{\varphi}_{\mathcal{H}}
    =\bra{\phi}\hat {\eta}
    \hat A\hat {\eta}^{-1}\ket{\varphi}_{\mathcal{H}},
    \label{eqn:O5-6O}
\EE
where $\phi, \varphi\in\Phi$. (For the derivation of (\ref{eqn:O5-6O}), see Appendix \ref{sec:A-2-4}.)
Considering that the relation (\ref{eqn:O5-6O}) is satisfied, $\hat{A}$ is identified with the $\eta$-quasi Hermitian operator for the bra-ket space.
%

\subsection{Similarity Transformation}

We suppose that there exists a similarity transformation concerning the non-Hermitian Hamiltonian $\hat H$ in the bra-ket space,
\BE
    \bra{\phi}{\hat h_\rho}\ket{\varphi}_{\mathcal{H}}
    :=\bra{\phi}\hat {\rho}
    \hat H \hat {\rho}^{-1}\ket{\varphi}_{\mathcal{H}},
    \label{eqn:O5-6}
\EE
for $\phi, \varphi\in\Phi$, where the transformed Hamiltonian $\hat h_\rho$ is Hermitian with respect to the bra-ket space:
$
\bra{\phi}{\hat h_\rho^\dagger}\ket{\varphi}_{\mathcal{H}}=
\bra{\phi}{\hat h_\rho}\ket{\varphi}_{\mathcal{H}}
$.
From the Hermitian property of $\hat h_\rho$,
we obtain
\BE
     \bra{\phi}\hat H^\dagger \ket{\varphi}_{\mathcal{H}} =
     \bra{\phi} \hat \rho^2 \hat H \hat \rho^{-2} \ket{\varphi}_{\mathcal{H}},
\EE
and therefore,
$
\hat \eta= \hat \rho^2
$ if and only if $\hat H$ is $\hat {\eta}$-quasi Hermitian.
The existence of the transformation (\ref{eqn:O5-6}) means that Hermitian quantum mechanics with $\hat h_\rho$ equivalent to the non-Hermitian ones with $\hat H$ can be constructed.

Observables in the Hermitian quantum mechanics are well-known, such as Hermitian operators.
Therefore, the observables $\hat {O}$ on the system described by $\hat H$ can be obtained by
\BE
    \bra{\phi}\hat {O}\ket{\varphi}_{\mathcal{H}} :=
    \bra{\phi} \hat  \rho^{-1} \, \hat o  \, \hat \rho \ket{\varphi}_{\mathcal{H}},
\EE
where $\hat {o}$ is the observable in the Hermitian system described by $\hat h_\rho$.
The operator $\hat {O}$ defined above is an $\hat{\eta}$-quasi Hermitian (see Appendix \ref{sec:A-2-5})
\BE \label{eq:quasi-Hermitian_for_observables}
    \hat{O}^\dagger = \hat{\eta} \hat{O} \hat{\eta}^{-1}.
\EE

\section{$\mathcal{PT}$-symmetric non-Hermitian system}
\label{sec:4}
In this section, we treat quantum systems with $\mathcal{PT}$-symmetry as non-Hermitian quantum systems with concrete symmetries.
%
%
Hereafter, we assume that operators are defined within bra-ket space and also omit the bra-kets.

\subsection{General formulation}
The $\mathcal{PT}$-symmetric Hamiltonian is satisfied by
\BE
    [\hat H, \hat{\mathcal{P}} \hat{ \mathcal{T}}] = 0,
\EE
where the parity and time transformation are defined by
$\hat{\mathcal{P}}:\, x\rightarrow -x, \, p\rightarrow -p$
and
$\hat{\mathcal{T}}:\, x\rightarrow x, \, p\rightarrow -p$, respectively.
Because this Hamiltonian is non-Hermitian, the eigenvectors of $\hat H$ are not orthogonal to each other under the conventional inner product
\BE
    \langle f | g \rangle := \int_\mathbb{R} dx f^*(x) g(x).
\EE
Therefore, to orthogonalize the eigenvectors, we have to introduce the inner product with an appropriate metric.
For example, under the inner product with the metric $\hat \eta=\hat{\mathcal{P}}$,
\BE
    \langle f | g \rangle_\mathcal{P} :=
    \langle f | \mathcal{P} g \rangle =
    \int dx f^*(x) g(-x),
\EE
the $\mathcal{PT}$-symmetric Hamiltonian behaves like Hermitian in this metric space.
However, this inner product derives $\langle f | f \rangle_\mathcal{P}=\pm$, which means the metric $\hat\eta=\hat{\mathcal{P}}$ is indefinite, and the Hilbert space based on this inner product is ill-defined.
To overcome this problem, we introduce an operator $\hat{\mathcal{C}}$, which flips the negative sign of the norm to positive, and define the following inner product with the metric $\hat \eta=\hat{\mathcal{P}}\hat{\mathcal{C}}$, such that
\BE
    \langle f | g \rangle_{\mathcal{PC}} :=
    \langle f | \mathcal{\hat P \hat C} g \rangle
\EE
called the $\mathcal{CPT}$-inner product in Ref.~\cite{Bender2002}.
The metric $\hat \eta=\hat{\mathcal{P}}\hat{\mathcal{C}}$ has the Hermitian property
\BE
    (\hat{\mathcal{P}}\hat{\mathcal{C}})^{\dagger} =
     \hat{\mathcal{T}} 
     (\hat{\mathcal{P}}\hat{\mathcal{C}})^T
     \hat{\mathcal{T}} =
     \hat{\mathcal{T}} \hat{\mathcal{C}}^T \hat{\mathcal{P}}^T \hat{\mathcal{T}} =
     \hat{\mathcal{T}} \hat{\mathcal{C}} \hat{\mathcal{P}} \hat{\mathcal{T}} =
     \hat{\mathcal{P}}\hat{\mathcal{C}},
\EE
from $[\hat{\mathcal{C}}, \hat{\mathcal{P}}\hat{\mathcal{T}}] = 0$.
Therefore, the metric $\hat{\mathcal{P}}\hat{\mathcal{C}}$ is a positive-definite invertible Hermitian operator.

From the definition of $\hat{\mathcal{C}}$, we obtain
$
    [\hat{\mathcal{C}}, \hat{\mathcal{P}}\hat{\mathcal{T}}] =
    [\hat{\mathcal{C}}, \hat{\mathcal{H}}] = 0,
$
and the Hamiltonian becomes
\BA
\hat{\mathcal{C}}\hat{\mathcal{P}}\hat{\mathcal{T}} \, \hat{H} \, (\hat{\mathcal{C}}\hat{\mathcal{P}}\hat{\mathcal{T}})^{-1} = \hat{H}, \\
\hat{\mathcal{P}}\hat{\mathcal{C}} \, \hat{H} \, (\hat{\mathcal{P}}\hat{\mathcal{C}})^{-1} = \hat{H}^\dagger,
\EA
for the $\mathcal{PC}$-transformation.
From these equations, we obtain
$
 \hat{\zeta} = \hat{\mathcal{C}}\hat{\mathcal{P}}\hat{\mathcal{T}} = (\hat{\mathcal{P}}\hat{\mathcal{C}})^{-1} \hat{\mathcal{T}} = \hat{\eta}^{-1} \hat{\mathcal{T}},
$
where $\hat{\zeta}$ is found in Sec. {\ref{sec:1}}.

We introduce the general $\mathcal{PC}$-pseudo-Hermitian operator $O$:
\BE
    \hat{O}^\dagger =
     \hat{\mathcal{P}}\hat{\mathcal{C}} \, \hat{O} \,  (\hat{\mathcal{P}}\hat{\mathcal{C}})^{-1}
\EE
that represents an observable of the system denoted by $\mathcal{H}$.
We note that this operator relates to the Hermitian operator $o$ in $L^2$-space $(L^2, \langle \cdot, \cdot \rangle)$:
\BE
    \hat {O} = (\hat{\mathcal{P}}\hat{\mathcal{C}})^{-\frac{1}{2}} \, \hat{o} \, (\hat{\mathcal{P}}\hat{\mathcal{C}})^{\frac{1}{2}}
\EE
e.g.
$
    \hat{X} = (\hat{\mathcal{P}}\hat{\mathcal{C}})^{-\frac{1}{2}} \, \hat{x} \, (\hat{\mathcal{P}}\hat{\mathcal{C}})^{\frac{1}{2}}
$
and
$
    \hat{P} = (\hat{\mathcal{P}}\hat{\mathcal{C}})^{-\frac{1}{2}} \, \hat{p} \, (\hat{\mathcal{P}}\hat{\mathcal{C}})^{\frac{1}{2}}
$.

Next, we construct the $\mathcal{PC}$-RHS.
We consider a $\mathcal{PT}$-pseudo Hermitian operator $O$: $\mathcal{D}(O) \rightarrow \mathcal{R}(O)$ for a positive invertible operator $\mathcal{P}{\mathcal{C}}$: $\mathcal{H} \rightarrow \mathcal{H}$.
We suppose that $O \Phi \subset \Phi$, ${\mathcal{P}}{\mathcal{C}} \Phi = \Phi$, and ${O}$ is continuous on $(\Phi, \tau_\Phi)$.
The $\mathcal{PC}$-RHS becomes the following triplet based on $\mathcal{H}_{\mathcal{PC}}$:
\BE
    \Phi \subset \mathcal{H}_{\mathcal{PC}} \subset \Phi', \quad
    \Phi \subset \mathcal{H}_{\mathcal{PC}} \subset \Phi^\times.
\EE
From the nuclear spectral theorem in ${\mathcal{PC}}$-RHS, even for the continuous spectrum, the spectral expansion can be obtained by
\BA
    \bra{\varphi}_{\mathcal{PC}} =
     \int_{Sp(\mathcal{PC})} \langle \varphi | \lambda' \rangle_{\mathcal{PC}} \bra{\lambda'}_{\mathcal{PC}} d\mu(\lambda'), \\
    \ket{\varphi}_{\mathcal{PC}} =
     \int_{{Sp(\mathcal{PC})}}
     \langle \lambda' | \varphi\rangle_{\mathcal{PC}} |\lambda'\rangle_{\mathcal{PC}} d\mu(\lambda')
\EA
where $\ket{\lambda'}_{\mathcal{PC}}$ and $\ket{\lambda}_{\mathcal{PC}}$ are generalized eigenvectors of $\hat{\eta}=\hat{\mathcal{P}}\hat{\mathcal{C}}$ and $\hat{H}$ (or $\hat{\mathcal{CPT}}$) defined by
\BA
    \hat{\mathcal{P}}\hat{\mathcal{C}} \ket{\lambda'} &=&
    \lambda' \ket{\lambda'}, \\
    \hat{H} \ket{\lambda}_{\mathcal{PC}} &=&
    \lambda_h \ket{\lambda}_{\mathcal{PC}}, \\
    \hat{\mathcal{C}}\hat{\mathcal{P}}\hat{\mathcal{T}} \ket{\lambda}_{\mathcal{PC}} &=&
    \lambda_{\mathcal{CPT}} \ket{\lambda}_{\mathcal{PC}}.
\EA

\subsection{Non-Hermitian $\mathcal{PT}$-symmetric oscillator}
Practically, it is necessary to obtain the representation of the metric operator and show the spectral expansion for a specific model.
In this subsection, we review this process throughout the simple non-Hermitian oscillator \cite{Swanson2004, Musumba2007, Quesne2007},
\BE \label{def:swanson}
    \hat H = \hbar\omega\left( \hat a^\dagger \hat a + \frac{1}{2} \right) + \hbar\alpha \hat a^2 + \hbar\beta \hat a^{\dagger, 2}
\EE
where $\omega,\,\alpha,\,\beta \in \mathbb{R}$, such that $\alpha\neq\beta$ and $\hat a,\, \hat a^\dagger$, are ladder operators that are represented as
\BE \label{eqn:a_to_xp}
    \hat a =
    \frac{1}{\sqrt{2}}\left( \frac{\hat x}{\ell} + i\frac{\ell}{\hbar}\hat p \right), \quad
    \hat a^\dagger = \frac{1}{\sqrt{2}}\left( \frac{\hat x}{\ell} - i\frac{\ell}{\hbar}\hat p \right),
\EE
where $\ell=\sqrt{\hbar/m\omega}$ is the typical length for the ordinary harmonic oscillator with the frequency $\omega$ and mass $m$.
By substituting (\ref{eqn:a_to_xp}) in (\ref{def:swanson}), one can see that the Hamiltonian has $\mathcal{PT}$-symmetry.
Therefore, the non-Hermitian Hamiltonian becomes Hermitian in $ \eta=\mathcal{PC}$ metric space.

Hereafter, we review the process to obtain the specific form of the metric $\hat \eta$.
If there is an operator $\rho$ that transforms the Hamiltonian to Hermitian $\hat h_\rho = \hat \rho \hat H \hat \rho^{-1}$ in $L^2$-space, $\eta=\rho^2$ becomes the metric.
We consider the specific form
\BE
    \hat \rho = \exp \hat A, \quad \hat A = \chi \hat a^\dagger \hat a + \kappa \hat a^2 + \kappa^* \hat a^{\dagger,2}
\EE
where $\chi^2 - 4 |\kappa|^2>0$.
From this transformation, we obtain
\BE
    \hat \rho \hat H \hat \rho^{-1} =
    U(\chi,\kappa) \left( \hat a^\dagger \hat a + \frac{1}{2} \right)
    + V(\chi, \kappa) \hat a^2 + W(\chi, \kappa) \hat a^{\dagger,2}.
\EE
The specific forms of the coefficients $U$, $V$, and $W$ are given in Ref.~\cite{Quesne2007}.
If $U\in\mathbb{R}$ and $W^*=V$ are satisfied, this converted Hamiltonian becomes Hermitian in $L^2$-space.
Therefore, the parameters are restricted by
\BE
    \kappa^* = \kappa, \quad
    \frac{\tanh 2\theta}{\theta} = \frac{\alpha-\beta}{(\alpha+\beta)\chi - 2\omega\kappa},
\EE
where
$\theta = \sqrt{\chi^2-4\kappa^2}$.
The Hermitian Hamiltonian 
{$\hat h_\rho$} is given by the following ordinary oscillator form:
\BE
    \hat h_\rho = \frac{1}{2M(z)} \hat p^2 + \frac{1}{2} M(z)\Omega^2 \hat{x}^2,
\EE
where
$z:=2\kappa/\chi$ with $z\in[-1, 1]$
is the free parameter,
$\Omega^2 \, = \,\omega^2 - 4\alpha\beta$, and
\BE
    M^{-1}(z) = \frac{-z(\alpha+\beta)+\omega-(\alpha+\beta-z\omega)\sqrt{1-\frac{(1-z^2)(\alpha-\beta)^2}{(\alpha+\beta-z\omega)^2}}}{(1+z)\hbar \ell^{-2}}
\EE
This Hamiltonian 
{$\hat h_\rho$} is the ordinary harmonic oscillator with the mass $M(z)$ and the strength of frequency $\Omega$ if $M(z)>0$ and $\Omega ^2>0$.
Hereafter, we consider this condition.
The eigenequation of the Hamiltonian $\hat h_\rho$ is given by
\BE
    \hat h_\rho u_n(x) = E_n u_n(x)
\EE
and the solutions are obtained by
\BA
    E_n &=& \hbar \Omega \left( n+\frac{1}{2} \right), \\
    u_n(x) &=& \sqrt{\frac{2^{-n}}{\sqrt{\pi}n! \mathrm{L}}} e^{-\frac{x^2}{2\mathrm{L}^2}} \textrm{H}_n \left(\frac{x}{\mathrm{L}} \right),
\EA
where $\textrm{H}_n$ is Hermitian polynomials and $\mathrm{L}=\sqrt{\hbar/M\Omega}$.
The completeness and orthogonality relations are given by
\BA
    \sum_{n=0}^\infty u_n(x) u_n(x')  =  \delta(x-x'),\label{eq:CompletenessRelation_NHO} \\
    \left\langle u_n, u_m \right\rangle_{L^2} 
     = \int^\infty_{-\infty} \!dx\, u_n(x) u_m(x)
     = \delta_{nm}. \label{eq:OrthogonalityRelation_NHO}
\EA

We obtain the observables in the non-Hermitian system, such that
\BA
    \hat {X}_\rho = \hat \rho^{-1} \hat{x} \hat \rho =
    \cosh \theta \cdot \hat{x} + i\frac{\ell^2}{\hbar}\frac{\chi-2\kappa}{\theta} \sinh\theta \cdot \hat{p}\\
    \hat {P}_\rho = \hat \rho^{-1} \hat {p} \hat \rho =
    \cosh \theta \cdot \hat {p} + i \frac{\hbar}{\ell^2} \frac{\chi+2\kappa}{\theta} \sinh\theta \cdot \hat {x},
\EA
and
\BE
    \hat H = \hat \rho^{-1} \hat h_\rho \hat \rho =
    \frac{1}{2M(z)} \hat{P}_\rho^2 + \frac{M(z)}{2} \Omega^2 \hat{X}_\rho^2
\EE

Here, we take $z=1$ and obtain
\BE
 M(z=1) = \frac{\hbar}{(\omega-\alpha-\beta) \ell^2},
\EE
and
\BE \label{eq:IV-B_metrix}
    \eta(x) = \rho^2(x)
=\exp \left(
        -\frac{\alpha-\beta}{\omega-\alpha-\beta} \frac{{x}^2}{\ell^2}
        \right).
\EE
%
We note that the transformation $\rho$ becomes bounded if $(\alpha-\beta)/(\omega-\alpha-\beta)>0$.
%
The observables 
in the non-Hermitian system
are given by
\BE
    \hat{X}_\rho=\hat{x}, \quad
    \hat{P}_\rho = \hat{p}+i\hbar\frac{\alpha-\beta}{(\omega-\alpha-\beta)\ell^2} \hat{x}.
\EE
%
We note that they are quasi-Hermitian.
(The proofs are denoted in Appendix \ref{sec:A-2-6}.)
\BE \label{eqn:IV-B_quasi-Hermitian}
    \eta(x) \hat {X}_\rho \eta^{-1}({x}) = \hat{X}_\rho^\dagger, \quad 
    \eta(x) \hat{P}_\rho \eta^{-1}(x) = \hat{P}_\rho^\dagger.
\EE
The eigenequation of Hamiltonian is shown by
\BA \label{eq:IV-B_eigenequations}
    \hat{H} \mathcal{U}_n(x) &=& \mathcal{E}_n \mathcal{U}_n(x),
\EA
with
\BA
    \mathcal{E}_n = E_n, \quad \mathcal{U}_n(x) = \rho^{-1}(x) u_n(x).
\EA
From (\ref{eq:CompletenessRelation_NHO}) and (\ref{eq:OrthogonalityRelation_NHO}), the completeness and orthogonality relations for $\{ \mathcal{U}_n(x) \}$ are derived as follows
\BA
    \sum_{n=0}^\infty \mathcal{U}_n(x) \eta(x') \mathcal{U}_m(x')  =  \delta(x-x'),\label{eq:CompletenessRelation} \\
    \left\langle \mathcal{U}_n, \mathcal{U}_m \right\rangle_\eta 
     = \int^\infty_{-\infty} \!dx\, \mathcal{U}_n(x) \eta(x) \mathcal{U}_m(x)
     = \delta_{nm}. \label{eq:OrthogonalityRelation}
\EA
%

The RHS for the ordinary harmonic oscillator $h_\rho$ in $x$-representation is described by the triplet of the Schwartz space $\mathcal{S}(\mathbb{R})$\cite{Bohm1978},
\BE
    \mathcal{S}(\mathbb{R}) \subset L^2 \subset{S}'(\mathbb{R}),\,\mathcal{S}^\times(\mathbb{R}).
    \label{eqn:SchwartzRHS}
\EE
%
%
%
%
The inner product orthogonalizing the eigenfunctions is the $\eta$-inner product, and the $\eta$-RHS concerning the $\mathcal{PT}$-symmetric non-Hermitian system can be constructed based on the triplet (\ref{eqn:SchwartzRHS}) with $\langle \varphi, \psi \rangle_{\eta}=\langle \varphi, \eta\psi \rangle_{L^2}=\int dx \varphi^*(x) \eta(x) \psi(x)$ 
$( \varphi,\psi \in L^2)$ as shown in Section~\ref{sec:2}.
From (\ref{eq:IV-B_metrix}), $\eta$ is continuous on $\mathcal{S}(\mathbb{R})$ and $\eta(x)\phi(x)\in \mathcal{S}(\mathbb{R})$ for $\phi(x)\in \mathcal{S}(\mathbb{R})$.
Therefore, the $\eta$-RHS in this system is represented by 
\BE
    \mathcal{S}(\mathbb{R}) \subset \widetilde {L^2_\eta} \subset \mathcal{S}'(\mathbb{R}), \, \mathcal{S}^\times(\mathbb{R}).
\EE
We obtain the spectral expansion of $\hat H$ as follows:
\BA 
\varphi(x) &=&
    \sum_n \left\{ \int_{-\infty}^{\infty} \!dx'\ \left(\eta(x')\varphi(x')\right) \mathcal{U}_n(x') \right\} \mathcal{U}_n(x), \nn\\
\hat H \varphi(x) &=&
    \sum_n  E_n \left\{ \int_{-\infty}^{\infty} \!dx'\ \left(\eta(x')\varphi(x')\right)  \mathcal{U}_n(x') \right\} \mathcal{U}_n(x), \nn\\
\hat H^\dagger \varphi(x) &=&
    \sum_n  E_n \left\{ \int_{-\infty}^{\infty} \!dx'\ \left(\eta(x')\varphi(x')\right) \eta^{-1}(x')  \mathcal{U}_n(x') \right\}\nn\\ &&\hspace{4cm}\times \eta(x)\mathcal{U}_n(x),
    \label{eq:SpectralTheorem_in_NHHO}
\EA
for $\varphi(x) \in \mathcal{S}(\mathbb{R})$.
(The proof is given in Appendix \ref{sec:A-2-7}.)
We introduce the bra-ket notation
\BE
    \varphi(x)=\langle x | \varphi \rangle_{L^2}, \quad
    \varphi^*(x)= \langle \varphi | x \rangle_{L^2},
\EE
and define
\BE
    \ket{\varphi}_\eta=\ket{\eta\varphi}_{L^2}, 
    \quad 
    \bra{\varphi}_\eta = \bra{\eta\varphi}_{L^2}.
\EE
They are related in Hermitian conjugate $[\ket{\varphi}_{L^2}]^\dagger=\bra{\varphi}_{L^2}$ and $[\ket{\varphi}_{\eta}]^\dagger=\bra{\varphi}_{\eta}$.
Therefore, using $\eta$-RHS, we obtain
\BA
    \ket{\varphi}_{L^2} 
    &=& \sum_{n=0}^{\infty} \langle \mathcal{U}_n|_\eta \, \hat\eta^{-1} |\varphi\rangle_{L^2} \ket{\mathcal{U}_n}_\eta, \\
    \ket{H\varphi}_{L^2} 
    &=& \sum_{n=0}^{\infty} E_n \langle \mathcal{U}_n|_\eta |\varphi\rangle_{L^2} \, \hat\eta^{-1} \ket{\mathcal{U}_n}_\eta, 
    \\
    \ket{H^\dagger \varphi}_{L^2} 
    &=& \sum_{n=0}^{\infty} E_n \langle \mathcal{U}_n|_\eta \hat{\eta}^{-1}|\varphi\rangle_{L^2} \, \hat\eta^{-1} \ket{\mathcal{U}_n}_\eta.
\EA
The spectral expansions for bras are obtained by taking Hermitian conjugate. 
We note that they are consistent with (\ref{eqn:O3-5a})-(\ref{eqn:O3-5d}). 
%
%
%
%
%
\color{black}
We note that spectral expansions are also shown for parameter regions other than $m>0, \, \Omega^2>0$ in a recent paper\cite{Fernandez2022}.
The general theory in this situation is a topic for the future.

As the first step to study the mathematically underlying space for the non-Hermitian quantum system characterized in the framework of RHS,
the bra-ket formulations for the non-Hermitian system with symmetry are focused on in this study.
There are several developments based on the RHS treatment; for example a study on a non-positive non-Hermitian system, such as the pseudo-Hermitian system \cite{Mos2010}, and the treatment of the $\eta$-RHS in terms of the unbounded inverse, positive operator \cite{Antoine2013}.
These topics will be dealt with in the future.
%


\section{Conclusion}
\label{sec:5}

In this study, Dirac's bra-ket formalism is investigated for a non-Hermitian system with a symmetrical structure.
An RHS in terms of the positive-definite metirc induced from a positive operator $\eta$, $\eta$-RHS, is established.
With the aid of the nuclear spectral theorem for $\eta$-RHS, we show the spectral expansions for the bra and ket vectors by the generalized eigenvectors of an $\eta$-quasi Hermitian operator and the complete bi-orthogonal system for the bra-kets, which endows the transformation theory between the Hermitian and non-Hermitian systems.
Extensions of the non-Hermitian operators to the bra-ket space are formulated to consider the observables in such a space for non-Hermitian systems.
As an application of our method a $\mathcal{PT}$-symmetric non-Hermitian system, such as that of a non-Hermitian oscillator, is focused on.

\appendix

\numberwithin{equation}{section}
\makeatletter

\section{Notations}
\label{sec:6-1}

Here, we summarize the notations used in this article:
\begin{description}
    \item[(i)] $\bra{\phi}_\mathcal{H} \cdot \ket{\varphi}_\mathcal{H}
    = \braket{\phi}{\varphi}_{\mathcal{H}}
    = \langle \phi,\, \varphi\rangle_\mathcal{H}
    =\bra{\phi}_\mathcal{H}(\varphi)
   =\ket{\varphi}_\mathcal{H}(\phi).$
    \item[(ii)] $\braket{\phi}{\varphi}_{\eta}
    =\bra{\phi}{\hat \eta}\ket{\varphi}_{\mathcal{H}}
    =\bra{\phi}_\mathcal{H}\cdot \hat {\eta}\ket{\varphi}_\mathcal{H}\\
    =\bra{\phi}_{\mathcal{H}}\hat {\eta} \cdot\ket{\varphi}_\mathcal{H}
    =\bra{\phi}_\mathcal{H}\cdot \ket{\eta\varphi}_\mathcal{H}\\
    =\bra{\eta \phi}_\mathcal{H}\cdot \ket{\varphi}_\mathcal{H}
    =\bra{\phi}_\mathcal{H}\cdot \ket{\varphi}_\eta
    =\bra{\phi}_{\eta} \cdot\ket{\varphi}_\mathcal{H}$.
    \item[(iii)] $\bra{\lambda}_\eta \cdot \ket{\varphi}_\mathcal{H}
    = \bra{\lambda}_H \hat {\eta}\ket{\varphi}_\mathcal{H}
    =\bra{\lambda}_\mathcal{H} \cdot \ket{\varphi}_\eta \\
    \equiv \braket{\lambda}{\varphi}_{\eta}
    =\bra{\lambda}_\eta(\varphi)$,\\
    $\bra{\phi}_\mathcal{H} \cdot \ket{\lambda}_\eta
    = \bra{\phi}_\mathcal{H} \hat{\eta}\ket{\lambda}_\mathcal{H}
    =\bra{\phi}_\eta\cdot \ket{\lambda}_\mathcal{H}\\
    = \braket{\phi}{\lambda}_{\eta}
    =\ket{\lambda}_\eta(\phi)$.
    \item[(iv)] $\braket{\lambda}{\lambda^{\prime}}_{\eta}
    =\bra{\lambda}_\mathcal{H} \cdot \ket{\lambda^{\prime}}_\eta
    =\bra{\lambda}_{\eta} \cdot\ket{\lambda^{\prime}}_\mathcal{H}
    =\bra{\lambda}_\mathcal{H} \cdot \hat{\eta}\ket{\lambda^{\prime}}_\mathcal{H}\\
    =\bra{\lambda}_{\mathcal{H}}\hat {\eta} \cdot\ket{\lambda^{\prime}}_\mathcal{H}
    =\bra{\lambda}{\hat \eta}\ket{\lambda^{\prime}}_{\mathcal{H}}.$
\end{description}
The notation (iv) is constructed by replacing $\phi$ and $\varphi$ in the notation (ii) with $\lambda$ and $\lambda^{\prime}$, respectively,

\section{Supplement for derivations of the relations}
\label{sec:A-2}

\subsection{The relation (\ref{eqn:O2-9})}
\label{sec:A-2-1}

As an $\eta$ is Hermitian with respect to $\langle \cdot, \cdot \rangle_\mathcal{H}$, we have
\BA
    \ket{\varphi}_\eta(\phi)
    =\langle \phi, \varphi \rangle_\eta
    =\langle \phi, \eta\varphi \rangle_\mathcal{H}
    =\langle \eta\phi, \varphi \rangle_\mathcal{H} \nn\\
    =\ket{\varphi}_\mathcal{H}(\eta\phi)
    =\hat {\eta}(\ket{\varphi}_\mathcal{H})(\phi)
    \label{eqn:A-2-1}
\EA
for any $\phi\in \Phi$.
Similarly,
\BA
    \bra{\varphi}_\eta(\phi)
    =\langle \varphi, \phi \rangle_\eta
    =\langle \varphi, \eta\phi \rangle_\mathcal{H} \nn\\
    =\bra{\varphi}_\mathcal{H}(\eta\phi)
    =\hat{\eta}(\bra{\varphi}_\mathcal{H})(\phi)
    \label{eqn:A-2-2}
\EA
for any $\phi\in \Phi$.
Therefore, the relation (\ref{eqn:O2-9}) is obtained.

\subsection{The relation (\ref{eqn:O3-5a})}
\label{sec:A-2-2}

Because $\eta^{-1}$, which is the inverse of an Hermitian operator $\eta$ with respect to $\langle \phi, \, \varphi \rangle_\eta$, is Hermitian, from (\ref{eqn:O2-10}) and (\ref{eqn:O3-2a}), we obtain
\BA
    \langle \phi, \, \varphi \rangle_\mathcal{H}
    & = & \langle \phi, \, \eta^{-1}\varphi \rangle_\eta
      = \langle \eta^{-1}\phi, \, \varphi \rangle_\eta \nn \\
    & = & \int_\mathbb{R}\braket{\phi}{\lambda}_{\eta} \braket{\lambda}{\eta^{-1}\varphi}_{\eta} d\mu (\lambda)  \nn \\
   & = & \int_\mathbb{R}\braket{\eta^{-1}\phi}{\lambda}_{\eta} \braket{\lambda}{\varphi}_{\eta} d\mu (\lambda) \nn \\
    & = & \int_\mathbb{R}\braket{\phi}{\lambda}_{\eta}\hat{\eta}^{-1}(\bra{\lambda}_{\eta})(\varphi)  d\mu (\lambda) \nn \\
   & = & \int_\mathbb{R}\hat{\eta}^{-1}\ket{\lambda}_{\eta}(\phi) \braket{\lambda}{\varphi}_{\eta} d\mu (\lambda),\nn\\
    \label{eqn:O3-40}
\EA
for any $\phi,\, \varphi \in \Phi$.
Therefore, the spectral expansions (\ref{eqn:O3-5a}) are obtained.

\subsection{The relation (\ref{eqn:O3-4-5})}
\label{sec:A-2-5}

Considering the complex conjugate of (\ref{eqn:O3-4-3a}),
\BA
    \braket{\omega}{\varphi}_\mathcal{H} ^*
    & = & \bra{\varphi}_\mathcal{H}\cdot \ket{\omega}_\mathcal{H} \nn \\
    & = & \int_\mathbb{R} \bra{\varphi}_\mathcal{H} \cdot \ket{\lambda}_\mathcal{H} \bra{\lambda}_\eta\cdot\ket{\omega}_\mathcal{H}  d\mu (\lambda)\nn \\
    & = & \int_\mathbb{R} \braket{\lambda}{\omega}_\eta\braket{\lambda}{\varphi}_\mathcal{H} ^*  d\mu (\lambda).
    \label{eqn:A-2-5-1}
\EA
Thus, we obtain
\BA
    \braket{\omega}{\varphi}_\mathcal{H}
    & = &  
    \bra{\varphi}_\mathcal{H} \cdot \ket{\omega}_\mathcal{H} ^* \nn\\
    & = &
    \left( 
    \int_\mathbb{R} \braket{\lambda}{\omega}_\eta\braket{\lambda}{\varphi}_\mathcal{H}^*  d\mu (\lambda)
    \right)^*
    \nn\\
    & = &  \int_\mathbb{R} \braket{\lambda}{\omega}_\eta^* \braket{\lambda}{\varphi}_\mathcal{H}  d\mu (\lambda).
    \label{eqn:A-2-5-2}
\EA
(\ref{eqn:A-2-5-2}) defines $\braket{\lambda}{\omega}_\eta ^*$, which is the complex conjugate of the transformation factor $\braket{\lambda}{\omega}_\eta$.
Comparing it to (\ref{eqn:O3-4-3a}), we obtain the relation $\braket{\lambda}{\omega}_\eta^*=\braket{\omega}{\lambda}_\eta$.
Similarly, we can show $\braket{\lambda}{\omega}_\mathcal{H}^*=\braket{\omega}{\lambda}_\mathcal{H}$.

\subsection{The relation (\ref{eqn:O5-6O})}
\label{sec:A-2-4}

Let us denote $(FG)\bra{\phi}$ by $\bra{\phi} G F$ where $F$ and $G$ operate on $\Phi^{\prime}\cup\Phi^{\times}$;
 $(FG)\bra{\phi}(\varphi)=G(\bra{\phi})(F\varphi)=\bra{\phi}GF(\varphi)$ for any $\phi, \varphi\in \Phi$.
Then, we obtain
\BA
    \bra{\phi}_\mathcal{H} \hat {\eta}^{-1}\hat {A}^{\dagger}
    = \bra{A\eta^{-1}\phi}_\mathcal{H}
    \label{eqn:A-2-4-1a}
\EA
for any $\phi\in \Phi$.
For any $\phi, \varphi\in \Phi$, we have
$
\bra{\phi}_\mathcal{H} \hat {\eta}^{-1}\hat {A}^{\dagger}(\varphi)
=\hat {A}^{\dagger}\hat {\eta}^{-1}\bra{\phi}_\mathcal{H}(\varphi)
=\bra{\phi}_\mathcal{H} (\eta^{-1}A^{\dagger}\varphi)
=\langle \phi, \, \eta^{-1}A^{\dagger}\varphi\rangle_\mathcal{H}
=\langle A\eta^{-1}\phi, \, \varphi\rangle_\mathcal{H}.
$
It follows from (\ref{eqn:A-2-4-1a}) that
for any $\phi,\varphi \in \Phi$,
\BA
    \bra{\phi}\hat {\eta}^{-1}
    \hat A^{\dagger}\hat {\eta}\ket{\varphi}_{\mathcal{H}}
    &=&  \bra{\phi}_\mathcal{H} \hat {\eta}^{-1}
    \hat A^{\dagger} \cdot \hat {\eta}\ket{\varphi}_{\mathcal{H}} \nn\\
     &=&  \bra{A\eta^{-1}\phi}_\mathcal{H} \cdot\ket{\eta\varphi}_{\mathcal{H}} \nn\\
     &=&  \langle A\eta^{-1}\phi, \,  \eta\varphi\rangle_\mathcal{H} \nn\\
     &=&  \langle \phi, \, \eta^{-1}A^{\dagger} \eta\varphi\rangle_\mathcal{H} \nn\\
     &=&  \langle \phi, \, A \varphi\rangle_\mathcal{H} \nn\\
     &=&  \bra{\phi}_\mathcal{H} \cdot\ket{A\varphi}_{\mathcal{H}}\nn\\
     &=&  \bra{\phi}{\hat A}\ket{\varphi}_{\mathcal{H}}.
     \label{eqn:A-2-4-1}
\EA
Using (\ref{eqn:A-2-4-1}),
the relation (\ref{eqn:O5-6O}) is derived as follows:
\BA
    \bra{\phi} \hat{A}^{\dagger} \ket{\varphi}_{\mathcal{H}}
    & = & \bra{\phi}\hat {\eta}\hat {\eta}^{-1}
    \hat {A}^{\dagger}\hat {\eta}\hat {\eta}^{-1}\ket{\varphi}_{\mathcal{H}}\nn \\
    & = & \bra{\eta\phi}_\mathcal{H}\hat{\eta}^{-1}
    \hat {A}^{\dagger}\hat {\eta}\ket{\eta^{-1}\varphi}_{\mathcal{H}}\nn \\
    & = & \bra{\eta\phi}_\mathcal{H}
    \hat {A}\ket{\eta^{-1}\varphi}_{\mathcal{H}}\nn \\
    & = &     \bra{\phi}\hat {\eta}
    \hat {A} \hat {\eta}^{-1}\ket{\varphi}_{\mathcal{H}}.
     \label{eqn:A-2-4-2}
\EA

\subsection{The relation (\ref{eq:quasi-Hermitian_for_observables})}
\label{sec:A-2-5}
From $\hat\rho^\dagger=\hat\rho$ and $\hat \eta=\hat \rho^2$, the quasi-Hermitisity of observable is derived as follows:
\BA
    \hat{O}^\dagger
    &=& \left( \hat{\rho}^{-1} \, \hat{o} \, \hat{\rho} \right)^\dagger \nn\\
    &=& \hat{\rho}^\dagger \, \hat{o}^\dagger \, \hat{\rho}^{-1,\dagger} \nn \\
    &=& \hat{\rho} \, \hat{o} \, \hat{\rho}^{-1}  \nn \\
    &=& \hat{\rho}^2 \, (\hat{\rho}^{-1} \, \hat{o} \, \hat{\rho}) \,  \hat{\rho}^{-2}  \nn \\
    &=& \hat{\eta} \, \hat{O} \, \hat{\eta}^{-1}
\EA
\subsection{The relation (\ref{eqn:IV-B_quasi-Hermitian})}
\label{sec:A-2-6}
Using the $x$-representation of metrix (\ref{eq:IV-B_metrix}), we obtain
\BA
    \eta(x) \hat X_\rho \eta^{-1}(x) &=& 
    \eta(x) \hat x \eta^{-1}(x) \nn\\
    &=& \eta(x) x \eta^{-1}(x) \nn\\
    &=& x = \hat x = \hat x^\dagger = \hat X_\rho^\dagger, \\
    \eta(x) \hat P_\rho \eta^{-1}(x) 
    &=& \eta(x) \left[ \frac{\hbar}{i}\frac{d}{dx}
    +i\hbar\frac{\alpha-\beta}{(\omega-\alpha-\beta)\ell^2} x \right] \eta^{-1}(x) \nn\\
    &=& \frac{\hbar}{i}\frac{d}{dx} -2i\hbar \frac{\alpha-\beta}{(\omega-\alpha-\beta)\ell^2} x \nn\\
    &&\hspace{1cm}+i\hbar\frac{\alpha-\beta}{(\omega-\alpha-\beta) \ell^2} x \nn \\
    &=& p - i\hbar\frac{\alpha-\beta}{(\omega-\alpha-\beta) \ell^2} x = 
    P_\rho^\dagger.
\EA
\subsection{The relation (\ref{eq:SpectralTheorem_in_NHHO})}
\label{sec:A-2-7}
The spectral theorem is derived as follows: for any $\varphi(x)\in\mathcal{S}(\mathbb{R})$
\BAN
    \varphi(x) 
    &=& \int_{-\infty}^\infty \!dx'\,
      \varphi(x') \delta(x-x') \nn\\
    &=& \sum_{n=0}^\infty \int_{-\infty}^\infty \!dx'\, \varphi(x') \mathcal{U}_n(x) \eta(x') \mathcal{U}_n(x') \nn\\
    &=& \sum_{n=0}^\infty \left\{ \int_{-\infty}^\infty \!dx'\, \eta(x')\varphi(x') \mathcal{U}_n(x') \right\} \mathcal{U}_n(x).
\EAN
Using above result, we obtain
\BAN
    \hat H \varphi(x)
    &=& \sum_{n=0}^\infty \left\{ \int_{-\infty}^\infty \!dx'\, \eta(x')\varphi(x') \mathcal{U}_n(x') \right\} \hat H \mathcal{U}_n(x) \nn\\
    &=& \sum_{n=0}^\infty E_n \left\{ \int_{-\infty}^\infty \!dx'\, \eta(x')\varphi(x') \mathcal{U}_n(x') \right\} \mathcal{U}_n(x),
\EAN
and
\BAN
    \hat H^\dagger \varphi(x)
    &=& \eta(x) H \left( \eta^{-1}(x)\varphi(x) \right) \\
    &=& \eta(x)\sum_{n=0}^\infty E_n \left\{ \int_{-\infty}^\infty \!dx'\, \eta^{-1}(x)\varphi(x) \eta(x') \mathcal{U}_n(x') \right\} \mathcal{U}_n(x) \\
    &=& \sum_{n=0}^\infty E_n \left\{ \int_{-\infty}^\infty \!dx'\, \eta^{-1}(x)\varphi(x) \eta(x') \mathcal{U}_n(x') \right\} \eta(x)\mathcal{U}_n(x).
\EAN

\color{black}

%
\bigskip

\noindent
{\bf Acknowledgement}

The authors are grateful to
Prof. Y.~Yamazaki, Prof. T.~Yamamoto, Prof. Y.~Yamanaka, and Prof. Emeritus A.~Kitada at Waseda University and Prof. K.~Iida at Kochi University for their useful comments and encouragement.
This work was supported by the Sasakawa Scientific Research Grant from The Japan Science Society and JSPS KAKENHI Grant Number 22K13976.\\

\noindent
{\bf Data Availability}

Data sharing is not applicable to this article as no new data were created or analyzed in this study.

\bigskip


\end{document}